# THE STATIONARY PHASE CONDITION APPLICABILITY TO THE STUDY OF TUNNEL EFFECT.

Alex E. Bernardini[1]

Department of Cosmic Rays and Chronology, State University of Campinas,
PO Box 6165, SP 13083-970, Campinas, Brazil,
*alexeb@ifi.unicamp.br*

**Abstract.** Some recent theoretical studies have tended to employ analytically-continuous *gaussian*, or infinite-bandwidth step pulses to examine tunneling process. The stationary phase method is often employed to this aim. However, *gaussian* functions do not have a well-defined front, such that their speed of propagation becomes ambiguous. Also, infinite bandwidth signals cannot propagate through any real physical medium (whose transfer function is therefore finite) without pulse distortion, which also leads to ambiguities in determining propagation velocity during the tunneling process. In this manuscript, we appoint some incompatibilities with the application of the method of stationary phase in deriving tunneling times which, eventually, can ruin the *superluminal* interpretation of transition times.

The study presented in this manuscript was stimulated by some previous theoretical results on *superluminality* upon the tunneling phenomenon [1–3]. When the time of passage $t_T$ of a quantum particle in a tunneling collision with a potential square barrier $V_0$ becomes essentially independent of the barrier extension $L$, it could be suggestive the possibility of a *superluminal* transmission process where the Hartman effect [4] takes place. Such a *superluminal* transit time is calculated by the classical principle of the stationary phase [5] where a narrow momentum distribution $g(k-k_0)$ centered around $k_0$ is used to construct a propagating wave packet. The stationary phase method can be successfully applied to describe the movement of the peak of the wave packet by assuming the phase which characterizes the propagation varies sufficiently smoothly around the maximum of $g(k - k_0)$ which *can* eventually be given by $k_0$ (stationary phase condition).

In particular, with regard to the tunneling processes, the method of stationary phase is commonly applied to find the position of the wave packet which goes through the potential barrier. The transmitted wave calculated second the Schroedinger formalism is given by [6]

$$\psi^T(x,t) = \int_0^w \frac{dk}{2\pi}\, g(k-k_0)\, T(k,L)\, \exp\left[i\, k\,(x-L) - i\,\frac{k^2}{2\,m}\,t + i\,\Theta(k,L)\right] \tag{1}$$

where

$$T(k,L) = \left\{1 + \frac{w^4}{4\,k^2\,(w^2-k^2)}\sinh^2\left[(w^2-k^2)^{\frac{1}{2}}\,L\right]\right\}^{-\frac{1}{2}}, \tag{2}$$

$$\Theta(k,L) = \arctan\left\{\frac{2\,k^2-w^2}{k\,(w^2-k^2)^{\frac{1}{2}}}\tanh\left[(w^2-k^2)^{\frac{1}{2}}\,L\right]\right\}, \tag{3}$$

and we have adopted $w = (2\,m\,V_0)^{\frac{1}{2}}$ and $\hbar = 1$. When the stationary phase condition is applied to the phase of Eq. (2) we obtain

$$\frac{d}{dk}\left\{k\,(x-L) - \frac{k^2}{2\,m}\,t + \Theta(k,L)\right\}\bigg|_{k=k_{\max}} = x - L - \frac{k_{\max}}{m}\,t + \frac{d\Theta(k,L)}{dk}\bigg|_{k=k_{\max}} = 0. \qquad (4)$$

The above result is frequently used to calculate the transit time $t_T$ for the transmitted wave packet when its peak appears at $x = L$,

$$t_T = \frac{d\Theta(k,L)}{dk}\bigg|_{k=k_{\max}} = \frac{2\,m}{k_{\max}\,\alpha}\left\{\frac{w^4\,\sinh[\alpha]\cosh[\alpha] - (2\,k_{\max}^2 - w^2)\,k_{\max}^2\,\alpha}{4\,k_{\max}^2\,(w^2 - k_{\max}^2) + w^4\,\sinh^2[\alpha]}\right\} \qquad (5)$$

where we have introduced the parameter $\alpha = \left(w^2 - k_{\max}^2\right)^{\frac{1}{2}} L$. The concept of *opaque* limit is introduced when we assume that $k_{\max}$ is independent of $L$ and then we make $\alpha$ tends to $\infty$ [1]. In this case, the transit time can be rewritten as

$$t_T^{OL} = \frac{2\,m}{k_{\max}\,\alpha}. \qquad (6)$$

In the literature, the value of $k_{\max}$ is frequently approximated by $k_0$, the maximum of $g(k-k_0)$, which, in fact, does not depend on $L$ and could lead us to interpret the Eq. (6) as a *superluminal* transmission time [1,3,7]. To clear up this point, we notice that when we take the *opaque* limit with $L$ going to $\infty$ and $w$ fixed as well as with $w$ going to $\infty$ and $L$ fixed, with $k_0 < w$ in both cases, the expression (6) suggests times corresponding to a transmission process performed with velocities higher than $c$ [1]. Such a *superluminal* interpretation was extended to the study of quantum tunneling through two successive barriers separated by a free region where it was theoretically verified that the total traversal time does not depend not only on the barrier widths, but also on the free region extension [3]. Besides, in a subsequent analysis, the same technique was applied to a problem with multiple successive barrier where the tunneling process was designated as a highly non-local phenomenon [7].

It would be perfectly acceptable to consider $k_{\max} = k_0$ for the application of the stationary phase condition when the momentum distribution $g(k - k_0)$ centered at $k_0$ is not modified by any bound condition. This is the case of the incydent wave packet before colliding with the potential barrier. Our criticism concern, however, with the way of obtaining all the above results for the the transmitted wave packet It does not have taken into account the bounds and enhancements imposed by the analytical form of the transmission coefficient. Hartman himself had already appointed that the

" Discussion of the time dependent behavior of a wave packet is complicated by its diffuse or spreading nature; however, the position of the peak of a *symmetrical* packet can be described with some precision"

To perform the correct analysis, we should calculate the right value of $k_{\max}$ to be substituted in Eq. (5) before taking the *opaque* limit. We are obliged to consider the relevant amplitude for the transmitted wave as the product of a symmetric momentum distribution $g(k - k_0)$ which describes the *incoming* wave packet by the modulus of the transmission amplitude $T(k,L)$ which is a crescent function of $k$. The maximum of this product representing the transmission modulating function would be given by the solution of the equation

$$g'(k - k_0)\,|T(k,L)| + g(k - k_0)\,|T(k,L)|' = g(k - k_0)\,|T(k,L)|\left[\frac{g'(k - k_0)}{g(k - k_0)} + \frac{|T(k,L)|'}{|T(k,L)|}\right] = 0 \quad (7)$$

Obviously, the peak of the modified momentum distribution will be shifted to the right of $k_0$ so that $k_{\max}$ will computed in the interval $]k_0, w[$. Moreover, we confirm that $k_{\max}$ presents an implicit dependence on $L$ as we can demonstrate by the numerical results presented in table 1 where we have found the maximum of $g(k - k_0)\,|T(k,L)|$ by assuming $g(k - k_0)$ is a *gaussian* function almost completely comprised in the interval $[0, w]$ given by

$$g(k - k_0) = \left(\frac{a^2}{2\,\pi}\right)^{\frac{1}{4}} \exp\left[-\frac{a^2(k - k_0)^2}{4}\right]. \qquad (8)$$

By increasing the value of $L$ with respect to $a$, the value of $k_{\max}$ obtained from the numerical calculations to be substituted in Eq. (5) also increases until $L$ reaches certain values for which the modified momentum distribution becomes unavoidably distorted. In this case, the relevant values of $k$ are concentrated around the upper bound value $w$. We shall show in the following that the value of $L$ which starts to distort the momentum distribution can be analytically obtained in terms of $a$. Now, if we take the *opaque* limit of $\alpha$ by fixing $L$ and increasing $w$, the above results immediately ruin the *superluminal* interpretation upon the result of Eq. (5) since $t_T^{OL}$ tends to $\infty$ when $k_{\max}$ is substituted by $w$.

Otherwise, when $w$ is fixed and $L$ tends to $\infty$, the parameter $\alpha$ calculated at $k = w$ becomes indeterminate. The transit time $t_T$ still tends to $\infty$ but now it exhibits a peculiar dependence on $L$ which can be easily observed by defining the auxiliary function

$$G(\alpha) = \frac{\sinh[\alpha]\cosh[\alpha] - \alpha}{\sinh^2[\alpha]} \tag{9}$$

which allow us to write

$$t_T^\alpha = \frac{2\,m}{w\,\alpha}\,G(\alpha). \tag{10}$$

When $\alpha \gg 1$, the transmission time always assume infinite values with an asymptotic dependence on $\left(w^2 - k^2\right)^{-\frac{1}{2}}$,

$$t_T^\alpha \approx \frac{2\,m}{w\,(w^2 - k^2)^{\frac{1}{2}}} \to \infty. \tag{11}$$

Only when $\alpha$ tends to 0 we have an explicit linear dependence on $L$ given by

$$t_T^0 = \frac{2\,m\,L}{w}\,\lim_{\alpha \to 0}\left\{\frac{G(\alpha)}{\alpha}\right\} = \frac{4\,m\,L}{3\,w} \tag{12}$$

In addition to the above results, the transmitted wave must be carefully studied in terms of the rapport between the barrier extension $L$ and the wave packet width $a$. For very thin barriers, i. e. when $L$ is much smaller than $a$, the modified transmitted wave packet presents substantially the same form of the incident one. For thicker barriers, but yet with $L < a$, the peak of the *gaussian* wave packet modulated by the transmission coefficient is shifted to higher energy values, i. e. $k_{\max} > k_0$ increases with $L$. For very thick barriers, i. e. when $L > a$, we are able to observe that the form of the transmitted wave packet is badly distorted with the greatest contribution coming from the Fourier components corresponding to the energy $w$ just above the top of the barrier in a kind of *filter effect*. We observe that the quoted distortion starts to appear when the modulated momentum distribution presents a *local maximal* point at $k = w$ which occurs when

$$\left.\frac{d}{dk}\left[g(k - k_0)\,|T(k,L)|\right]\right|_{k=w} > 0. \tag{13}$$

Since the derivative of the *gaussian* function $g(k - k_0)$ is negative at $k = w$, the Eq. (13) gives the relation

$$-\frac{g'(w - k_0)}{g(w - k_0)} < \lim_{k \to w}\left[\frac{T'(k,L)}{T(k,L)}\right] = \frac{wL^2}{4}\frac{\left(1 + \frac{wL^2}{3}\right)}{\left(1 + \frac{wL^2}{4}\right)} < \frac{wL^2}{3} \tag{14}$$

which effectively represents the inequality

$$\frac{a^2}{2}(w - k_0) < \frac{wL^2}{3} \quad \Rightarrow \quad L > \sqrt{\frac{3}{2}}\,a\left(1 - \frac{k_0}{w}\right). \tag{15}$$

Due to the *filter effect*, the amplitude of the transmitted wave is essentially composed by the plane wave components of the front tail of the *incoming* wave packet which reaches the first barrier interface before the peak arrival. Meanwhile, only whether we had *cut off* the momentum distribution at a value of $k$ smaller than $w$, i. e. $k \approx (1 - \delta)w$, the *superluminal* interpretation of the transition time (6) could be recovered. In this case, independently of the way as $\alpha$ tends to $\infty$ the value assumed by the transit time would be approximated by $t_T^\alpha \approx 2\,m/w\,\delta$ which is a finite quantity. Such a finite value would confirm the hypothesis of *superluminality*. However, the *cut off* of the momentum distribution at $k \approx (1-\delta)w$ increases the amplitude of the tail of the incident wave as we can observe in Fig.1. It contributes so relevantly as the peak of the incident wave to the final composition of the transmitted wave and creates an ambiguity in the definition of the *arrival* time.

To conclude, we are convinced that the use of a step-discontinuity to analyze signal transmission in tunneling processes deserves a more careful analysis than the immediate application of the stationary phase method. A suggestive possibility corresponds to the use of the multiple peak decomposition technique developed for the above barrier diffusion problem [8]. In this case, the problem is still completely "open" because the conservation of probabilities was not clearly understood. There have also been some tentative of yielding complex time delays, ultimately due to a complex propagation constant. This has caused some confusion, with Landauer denying the physical reality to an imaginary time [9]. In parallel to the most sensible candidate for tunneling times [10,11], a phase-space approach have been use to determine a semi-classical traversal time [12]. This semi-classical method makes use of complex trajectories which, by its turn, enables the definition of real traversal times in the complexified phase space. In summary, we agree with the assertion that it is necessary to continue to search for a general answer for the time aspects of the tunneling process.

## Acknowledgments

The authors thank the University of Lecce for the hospitality and the CAPES (A.E.B.) for Financial Support.

## References

1. J. Jakiel, V. S. Olkhovsky and E. Recami, *On superluminal motions in photon and particle tunnellings*, Phys. Lett. **A 248**, 156 (1998).
2. R. Y. Chiao, *Tunneling Times and Superluminality: a Tutorial*, quant-ph/9811019.
3. V. S. Olkhovsky, E. Recami and G. Salesi, *Superluminal effects for quantum tunneling through two successive barriers*, Europhys. Lett. **57**, 879 (2002).
4. T. E. Hartman, *Tunneling of a wave packet*, J. Appl. Phys. **33**, 3427 (1962).
5. E. P. Wigner, *Lower limit for the energy derivative of the scattering phase shift*, Phys. Rev. **98**, 145 (1955).
6. C. Cohen-Tannoudji, B. Diu and F. Laloë, *Quantum Mechanics* (John Wiley & Sons, Paris, 1977).
7. S. Esposito, *Multibarrier tunneling*, Phys. Rev. **E 67**, 016609 (2003).
8. A. E. Bernardini, S. De Leo and P. Rotelli, *Above Potential Barrier Diffusion*, quant-ph/0408028
9. R. Landauer, *Barrier Traversal Time*, Nature **341**, 567 (1989).
10. E. H. Hauge and J. A. Stvneng, *Tunnelling Times, a critical review*, Rev. Mod. Phys. **61**, 917, (1989).
11. R. Landauer and Th. Martin, *Barrier Interaction Time in Tunneling*, Rev. Mod. Phys. **66**, 217 (1994).
12. A. L. Xavier and M. A. M. de Aguiar, *Phase-space approach to the tunnel effect: a new semiclassical traversal time*, Phys. Rev. lett. **79**, 3323 (1997).

| $w\,a$ \ $L/a$ | 1.5 | 2.0 | 4.0 | 6.0 | 8.0 | 10 | 20 |
|---|---|---|---|---|---|---|---|
| 0.00 | 1.0000 | 1.0000 | 1.0000 | 1.0000 | 1.0000 | 1.0000 | 1.0000 |
| 0.05 | 1.0062 | 1.0188 | 1.1777 | 1.4156 | 1.6238 | 1.7726 | 1.9834 |
| 0.10 | 1.0235 | 1.0648 | 1.3799 | 1.6769 | 1.8547 | 1.9397 | 2.0051 |
| 0.15 | 1.0489 | 1.1223 | 1.5349 | 1.8251 | 1.9505 | 1.9937 | 2.0133 |
| 0.20 | 1.0794 | 1.1825 | 1.6571 | 1.9178 | 2.0000 | 2.0204 | 2.0203 |
| 0.25 | 1.1129 | 1.2420 | 1.7575 | 1.9813 | 2.0317 | 2.0390 | 2.0272 |
| 0.30 | 1.1478 | 1.3001 | 1.8430 | 2.0289 | 2.0562 | 2.0551 | 2.0342 |
| 0.35 | 1.1836 | 1.3565 | 1.9185 | 2.0679 | 2.0779 | 2.0704 | 2.0413 |
| 0.40 | 1.2196 | 1.4116 | 1.9874 | 2.1025 | 2.0986 | 2.0857 | 2.0484 |
| 0.45 | 1.2558 | 1.4657 | 2.0524 | 2.1350 | 2.1191 | 2.1012 | 2.0556 |
| 0.50 | 1.2921 | 1.5194 | 2.1155 | 2.1668 | 2.1399 | 2.1170 | 2.0628 |
| 0.55 | 1.3285 | 1.5729 | 2.1785 | 2.1988 | 2.1611 | 2.1331 | 2.0701 |
| 0.60 | 1.3649 | 1.6266 | 2.2429 | 2.2314 | 2.1828 | 2.1495 | 2.0775 |
| 0.65 | 1.4015 | 1.6809 | 2.3101 | 2.2651 | 2.2051 | 2.1663 | 2.0850 |
| 0.70 | 1.4383 | 1.7360 | 2.3819 | 2.3002 | 2.2281 | 2.1834 | 2.0925 |
| 0.75 | 1.4751 | 1.7920 | 2.4599 | 2.3367 | 2.2518 | 2.2009 | 2.1001 |
| 0.80 | * | 1.8489 | 2.5466 | 2.3751 | 2.2761 | 2.2188 | 2.1078 |
| 0.85 | * | 1.9065 | 2.6456 | 2.4154 | 2.3013 | 2.2371 | 2.1155 |
| 0.90 | * | 1.9646 | 2.7627 | 2.4578 | 2.3272 | 2.2558 | 2.1234 |
| 0.95 | * | * | 2.9091 | 2.5028 | 2.3540 | 2.2750 | 2.1313 |
| 1.00 | * | * | 3.1137 | 2.5504 | 2.3818 | 2.2947 | 2.1392 |

**Table 1.** The values of $k$ are numerically obtained by increasing the barrier extension $L$ which is given in terms of the wave packet width $a$ for different values of the potential barrier high expressed in terms of $w\,a$. We have fixed the incoming momentum by setting $k_0\,a$ equal to 1. For the values of $L$ marked with *, we can demonstrate by means of Eqs. (14-15) that the modulated momentum distribution has already been completely distorted and the maximum loss its meaning in the context of applicability of the method of stationary phase.

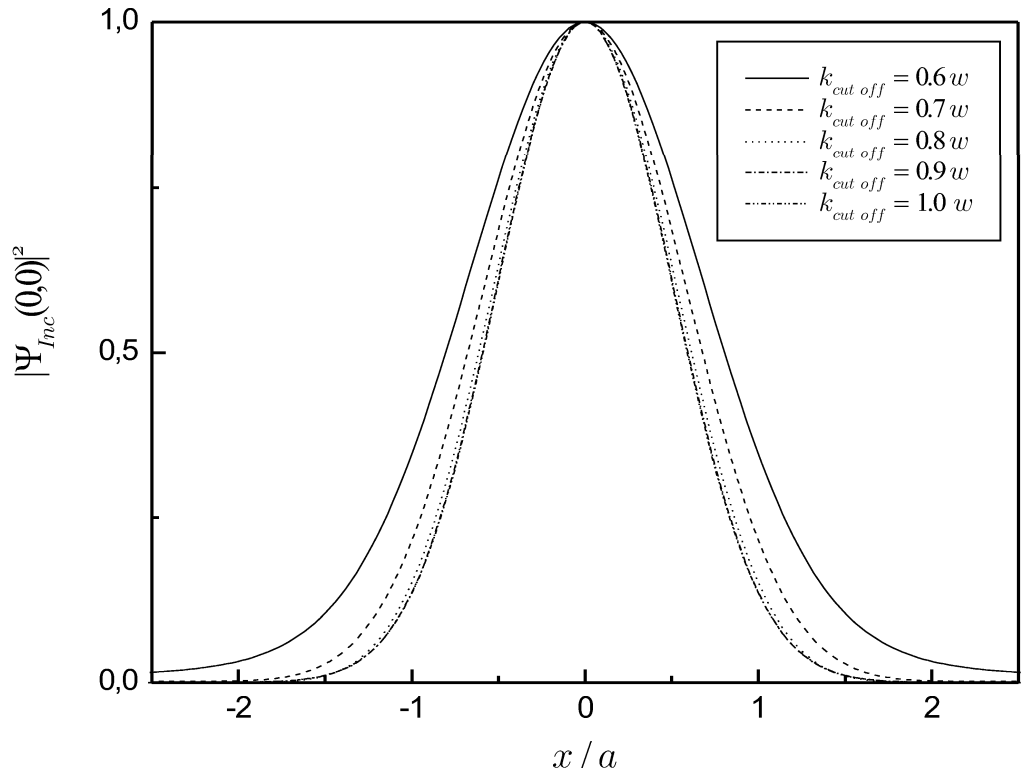

**Fig. 1.** Dependence of the wave packet shape on the *cut off* value of a momentum distribution centered around $k_0 = 0.5w$ with the values of $k$ comprised between 0 and $k_{cut\ off}$.